\documentclass[aps,twocolumn,showpacs, showkeys,nofootinbib,floatfix]{revtex4-1}

\usepackage{amsmath}
\usepackage{amsfonts}
\usepackage{txfonts}
\usepackage{graphicx}
\usepackage{dcolumn}
\usepackage{bbm}
\usepackage{amssymb}
\usepackage{latexsym}
\usepackage{titlesec}
\usepackage[colorlinks=true, linkcolor=red, citecolor=blue]{hyperref}
\usepackage{longtable}

\begin{document}

\title{FRCAMB: An $f(R)$ Code for Anisotropies in the Microwave Background}

\author{Lixin Xu$^{1,2}$}
\email{Corresponding author: lxxu@dlut.edu.cn}

\affiliation{$^{1}$Institute of Theoretical Physics, School of Physics \&
Optoelectronic Technology, Dalian University of Technology, Dalian,
116024, P. R. China}

\affiliation{$^{2}$State Key Laboratory of Theoretical Physics, Institute of Theoretical Physics, Chinese Academy of Sciences, Beijing, 100190, P. R. China}

\begin{abstract}
An $f(R)$ gravity model is proposed to realize a late time accelerated expansion of our Universe. To test the viability of an $f(R)$ gravity model through cosmic observations, the background evolution and the Einstein-Boltzmann equation should be solved for studying the effects on the cosmic microwave background power spectrum and on the matter power spectrum. In the market, we already have the modified versions of {\bf CAMB} code, for instance {\bf EFTCAMB} and {\bf MGCAMB}. However, in these publicly available Einstein-Boltzmann codes, a specific background cosmology, for example the $\Lambda$CDM or $w$CDM, is assumed. This assumption would be non-proper for a specific $f(R)$ model where the background evolution may be different from a $\Lambda$CDM cosmology. Therefore the main task for this paper is to present a code to calculate the anisotropies in the microwave background for any $f(R)$ gravity model based on {\bf CAMB} code, i.e. {\bf FRCAMB}, where the background and perturbation evolutions are included consistently. As results, one can treat {\bf FRCAMB} as a blackbox to output the CMB power spectrum and matter power spectrum, once an $f(R)$ function, its first two derivative with respect to $R$, i.e. $f_R\equiv df/dR$, $f_{RR}\equiv d^2f/dR^2$ and the reasonable values of the model parameters are inputted properly. As by-products, one can also output the effective equation of state of $f(R)$ model, the evolution of the dimensionless energy densities and other interesting cosmological quantities. 
\end{abstract}


\maketitle

\section{Introduction}

The late time accelerated expansion of our Universe demands a modification of general relativity (GR) at large scale or an addition of an extra exotic energy component, see the monograph \cite{ref:book} and references therein. Usually one believes that a modified gravity (MG) model is degenerate to a dark energy model at the background level. That is to say for a modified gravity model there exists a dark energy model having an effective equation of state (EoS) which produces the same expansion history as that of the modified gravity model. Thus for the same background evolution one expects to distinguish modified gravity from dark energy models through the dynamical evolutions of perturbations. In the literature, many efforts have been made aiming to detect the possible deviation to GR through a parameterized Poisson equation and the slip of the Newtonian potentials while fixing the background evolution to a $\Lambda$CDM cosmology. The parameterized formalism with a fixing $\Lambda$CDM background is proper, when one only concerns a general MG model which probably has no explicit Lagrangian in general. Thus the fact would be embarrassed even if a significant deviation from GR is confirmed, because the explicit MG theory is still unknown in this parameterized formalism. Of course before fixing a proper MG, a significant deviation to GR should be confirmed by cosmic observations. This is the main reason why this parameterized formalism is still plausible now. However unfortunately currently available cosmic observations have not found any significant deviation to GR \cite{ref:planck2015MG}.      

In this paper, instead of considering the parameterized formalism, we will mainly focus on a family of modified gravity, i.e. $f(R)$ gravity model, not only including the background evolution but also including the perturbation evolution. When an $f(R)$ gravity model is considered, it becomes non-proper to keep a $\Lambda$CDM background evolution. It is mainly because that a $\Lambda$CDM background evolution specifies the form of $f(R)$ \cite{ref:FRCAMB}. As well known, a reasonable Universe should mainly experience three important epochs consequently: radiation, matter and (effective) dark energy dominated stages. In terms of the effective EoS       
\begin{equation}
w_{eff}=-1-\frac{2}{3}\frac{H^{\prime}}{H},
\end{equation}
our Universe should vary from $w_{eff}=1/3$ (radiation dominated epoch), $w_{eff}=0$ (matter dominated epoch) to $w_{eff}<-1/3$ (dark energy dominated epoch) at different epochs. Therefore the background evolution provides a preliminary and valuable test for the viability of an $f(R)$ gravity model. The geometrical measurements from the luminosity of type Ia supernovae (SNe) as standard candle, the angular diameter distance of baryon acoustic oscillation (BAO) as standard ruler and the positions of the peaks of the cosmic microwave background (CMB) radiation power spectrum can constrain an $f(R)$ gravity model at the background level extensively.

In addition to the geometrical measurements, the dynamical measurements related to the evolution of the perturbation at the linear and non-linear scales break the degeneracy of model parameters and provide even stronger constraint to an $f(R)$ gravity model. To study the CMB power spectrum, one should study the perturbation evolution for an $f(R)$ gravity model, i.e. solve the full Einstein-Boltzmann equation for photons. In the market, we already have the Einstein-Boltzmann equation solvers for the modified gravity models (including $f(R)$ gravity model as a special case), for instances {\bf MGCAMB} \cite{ref:MGCAMB}, {\bf EFTCAMB} \cite{ref:EFTCAMB}  and {\bf FRCAMB} \cite{ref:FRCAMB} which are modified version of {\bf CAMB} package \cite{ref:CAMB}. However for the {\bf MGCAMB} code, a $\Lambda$CDM or $w$CDM background is assumed when the CMB power spectrum is shown in an $f(R)$ gravity model. Obviously it is non-proper once the background evolution for an $f(R)$ gravity model deviates from that of $\Lambda$CDM or $w$CDM. The situation is little different for the {\bf EFTCAMB} code, where a background evolution is fixed and an $f(R)$ model is reconstructed from this fixed expansion history.   
Although the background evolution is consistent to the perturbation evolution, the freedom of choosing different $f(R)$ gravity model is lost. Our {\bf FRCAMB} code is designed for any $f(R)$ gravity model at the background and perturbation evolutions based on the publicly available {\bf FRCAMB} code \cite{ref:FRCAMB}, once a form of $f(R)$ and its first two derivatives with respect to $R$ are inputted. 

This paper is structured as follows. In Section \ref{sec:FRBackground}, the background evolution equations for an $f(R)$ gravity model are presented. The perturbation evolution is given in Section \ref{sec:perturbation}. The CMB power spectrum and matter power spectrum are given in Section \ref{sec:powers}. Section \ref{sec:conclusion} is the conclusion.    

\section{Background Evolution for an $f(R)$ Gravity Model} \label{sec:FRBackground}

The Einstein-Hilbert action for an $f(R)$ gravity model reads as
\begin{equation}
S=\frac{1}{16\pi G}\int d^4x \sqrt{-g}\left[R+f(R)\right]+\int d^4x \sqrt{-g}(\mathcal{L}_{m}+\mathcal{L}_{r}),
\end{equation}
where $\mathcal{L}_{m}$ and $\mathcal{L}_{r}$ are the Lagrangian of matter and radiation respectively, which will not include the mysterious dark energy as  the late time accelerated expansion of our Universe can be realized by the proposed $f(R)$ gravity. For recent reviews for modified gravity theory, see Refs. \cite{ref:MGR1,ref:MGR2,ref:MGR3,ref:MGR4}. Doing variation with respect to the metric $g_{\mu\nu}$ for the Einstein-Hilbert action, one obtains a generalized Einstein equation which relates the geometry of space-time to the distribution of energy-momentum
\begin{equation}
G_{\mu\nu}+f_R R_{\mu\nu}-\left(\frac{1}{2}f-\Box f_R\right)g_{\mu\nu}-\nabla_{\mu}\nabla_{\nu}f_R=8\pi G T_{\mu\nu},\label{eq:frEinstein}
\end{equation}
where
\begin{equation}
f_R\equiv \frac{df(R)}{dR}
\end{equation}
plays as an effective scalar filed.
It is obvious that the general relativity is recovered when $f(R)=0$.  
The Friedmann equation for an $f(R)$ gravity reads as
 \begin{equation}
H^2+f_{R}\left(H^2-\frac{R}{6}\right)+\frac{f}{6}+H^2 f_{RR}R^{\prime}=\frac{\kappa^2}{3}\left(\rho_m+\rho_r\right),\label{eq:frFrid}
\end{equation} 
where $\kappa^2=8\pi G$, $f_{RR}\equiv d^2 f(R)/d R^2$, the prime $^{\prime}$ denotes the derivative with respect to the nature logarithm of the scale factor $a$, i.e. $\ln a$; $H=\dot{a}/a$ is the expansion rate of our Universe; and $\rho_{i},i=m,r$ is the energy density of the matter (cold dark matter+baryon) and radiation. Replacing the Ricci scalar $R$ by $R=12H^2+6HH^{\prime}$, the above equation (\ref{eq:frFrid}) can be recast to 
\begin{equation}
H^2-f_{R}(HH^{\prime}+H^2)+\frac{f}{6}+H^2f_{RR}R^{\prime}=m^2(a^{-3}+\frac{\Omega_{r}}{\Omega_{m}}a^{-4}),\label{eq:frFride2}
\end{equation}
where 
\begin{equation}
m^2=\frac{\kappa^2\rho_{m0}}{3}\simeq (8315\text{ Mpc})^{-2}\left(\frac{\Omega_m h^2}{0.13}\right),
\end{equation}
and $\Omega_{i}=\kappa^2\rho_{i0}/3H^2_0,i=m,r$ is the dimensionless energy density of the matter (cold dark matter+baryon) and radiation. The subscript $0$ denotes the corresponding value at present $a=1$. In similar to \cite{ref:HSmodel}, defining the dimensionless variables 
\begin{eqnarray}
y_{H}&=&\frac{H^2}{m^2}-a^{-3}-\frac{\Omega_{r}}{\Omega_{m}}a^{-4},\\
y_{R}&=&\frac{R}{m^2}-3a^{-3},
\end{eqnarray}
and using Eq. (\ref{eq:frFride2}), one obtains the differential equations for $\{y_H,y_R\}$
\begin{eqnarray}
y^{\prime}_{H}&=&\frac{1}{3}y_{R}-4y_{H},\\
y^{\prime}_{R}&=&9a^{-3}-\frac{1}{m^2 f_{RR}\left(y_{H}+a^{-3}+\frac{\Omega_{r}}{\Omega_{m}}a^{-4}\right)}\nonumber\\
&\times&\left[y_{H}-f_{R}\left(\frac{y_{R}}{6}-y_{H}-\frac{1}{2}a^{-3}-\frac{\Omega_{r}}{\Omega_{m}}a^{-4}\right)+\frac{f}{6m^2}\right].
\end{eqnarray}
The solutions of this equations describe the background evolution for an $f(R)$ gravity model, once the initial conditions at present $a_0=1$ are given
\begin{eqnarray}
\left.y_{H}\right|_{a=1}&=&\frac{H^2_0}{m^2}-1-\frac{\Omega_{r}}{\Omega_{m}},\\
\left.y_{R}\right|_{a=1}&=&6(1-q_0)\frac{H^2_0}{m^2}-3,
\end{eqnarray}
where we have used the relation $R_0=6H^2_0(1-q_0)$, here $q_0$ is the value of the deceleration parameter $q=-\ddot{a}a/\dot{a}^2$ at present. $q_0$ is a free model parameter which is related to $f_{R0}$ when an $f(R)$ gravity model is known. Therefore as a comparison to the literature, $f_{R0}$ is a derived model parameter.

The effective dark energy density and pressure for an $f(R)$ gravity are given by \cite{ref:book}
\begin{widetext}
\begin{eqnarray}
\kappa^2\rho_{DE}&=&\frac{1}{2}\left(f_{R}R-f\right)-3H^2f_{RR}R^{\prime}+3H^2\left(f_{R0}-f_{R}\right)\nonumber\\
&=&m^2\left[\frac{1}{2}\left(f_{R}\frac{R}{m^2}-\frac{f}{m^2}\right)-3\frac{H^2}{m^2}f_{RR}R^{\prime}+3\frac{H^2}{m^2}\left(f_{R0}-f_{R}\right)\right],\label{eq:rhode}\\
\kappa^2p_{DE}&=&f^{\prime\prime}_{RR}H^2+f_{RR}R^{\prime}HH^{\prime}+2H^2f^{\prime}_{R}-\frac{1}{2}\left(Rf_{R}-f\right)-\left(2HH^{\prime}+3H^2\right)\left(f_{R0}-f_{R}\right)\nonumber\\
&=&m^2\left\{\frac{H^2}{m^2}f^{\prime\prime}_{R}+\left[\frac{1}{2}\left(\frac{H^2}{m^2}\right)^{\prime}+2\frac{H^2}{m^2}\right]f_{RR}R^{\prime}-\frac{1}{2}\left(f_{R}\frac{R}{m^2}-\frac{f}{m^2}\right)-\left[\left(\frac{H^2}{m^2}\right)^{\prime}+3\frac{H^2}{m^2}\right]\left(f_{R0}-f_{R}\right)\right\}\label{eq:pde}.
\end{eqnarray}
\end{widetext} 
The EoS of the effective dark energy $w_{DE}=p_{DE}/\rho_{DE}$ can be obtained easily
\begin{eqnarray}
w_{DE}&=&-1+\frac{2\left(f^{\prime\prime}_{R}H^2+f^{\prime}_{R}H^{\prime}H\right)-2H^2f^{\prime}_{R}-4HH^{\prime}\left(f_{R0}-f_{R}\right)}{\left(Rf_{R}-f\right)-6H^2f^{\prime}_{R}+6H^2\left(f_{R0}-f_{R}\right)},\nonumber\\
&=&-1+\frac{2\frac{H^2}{m^2}f^{\prime\prime}_{RR}+\left[\left(\frac{H^2}{m^2}\right)^{\prime}-2\frac{H^2}{m^2}\right]f_{RR}R^{\prime}-2\left(\frac{H^2}{m^2}\right)^{\prime}\left(f_{R0}-f_{R}\right)}{f_{R}\frac{R}{m^2}-\frac{f}{m^2}-6\frac{H^2}{m^2}f_{RR}R^{\prime}+6\frac{H^2}{m^2}\left(f_{R0}-f_{R}\right)}.\label{eq:wde}
\end{eqnarray}
The effective EoS of our Universe becomes
\begin{equation}
w_{eff}=-1-\frac{2}{3}\frac{H^{\prime}}{H}=-1-\frac{1}{3}\frac{\left(\frac{H^2}{m^2}\right)^{\prime}}{\left(\frac{H^2}{m^2}\right)},\label{eq:weff}
\end{equation}
which enables us to test the viability of the $f(R)$ model at the background level. The main reason is that $w_{eff}$ should vary from $w_{eff}=1/3$ (radiation dominated epoch), $w_{eff}=0$ (matter dominated epoch) to $w_{eff}<-1/3$ (dark energy dominated epoch) at different epochs. With the help of the following expressions,   
\begin{eqnarray}
\frac{H^2}{m^2}f^{\prime\prime}_{R}&=&-f_{R}\left(\frac{H^2}{m^2}\right)^{\prime}-\left[\frac{1}{2}\left(\frac{H^2}{m^2}\right)^{\prime}-\frac{H^2}{m^2}\right]f_{RR}R^{\prime}-(3a^{-3}+4\frac{\Omega_{r}}{\Omega_{m}}a^{-4}),\\
f_{RR}R^{\prime}&=&m^2f_{RR}(y^{\prime}_{R}-9a^{-3}),\\
\left(\frac{H^2}{m^2}\right)^{\prime}&=&\frac{1}{3}y_{R}-4y_{H}-3a^{-3}-4\frac{\Omega_{r}}{\Omega_{m}}a^{-4},
\end{eqnarray}
finally the Eqs. (\ref{eq:rhode}), (\ref{eq:pde}), (\ref{eq:wde}) and (\ref{eq:weff}) can be rewritten as functions of the dimensionless variables $\{y_{H},y_{R}\}$. Until now, we have obtained the background evolution 
\begin{equation}
H^2=\frac{\kappa^2}{3(1+f_{R0})}\left(\rho_m+\rho_{r}+\rho_{DE}\right).
\end{equation}
for an $f(R)$ gravity and the effective energy density, pressure and EoS.

\begin{center}
\begin{figure}[tbh]
\includegraphics[width=9.25cm]{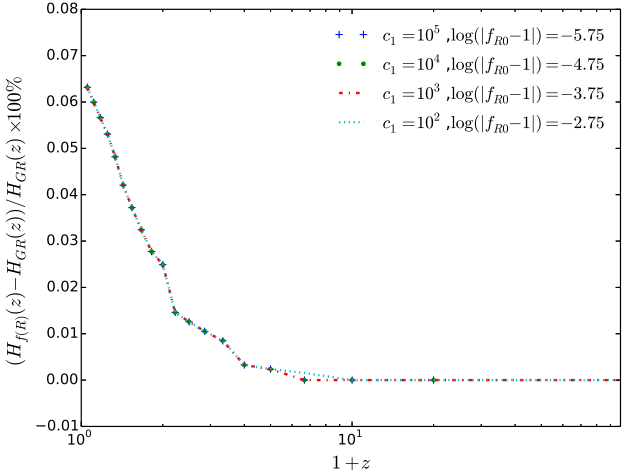}
\caption{The relative difference, $(H_{f(R)}(z)-H_{GR}(z))/H_{GR}(z)\times 100 \%$, of the Hubble parameters for the HS ($n=1$, $q_0=-0.65$) and $\Lambda$CDM model.}\label{fig:fRH}
\end{figure}
\end{center}

\begin{center}
\begin{figure}[tbh]
\includegraphics[width=9.25cm]{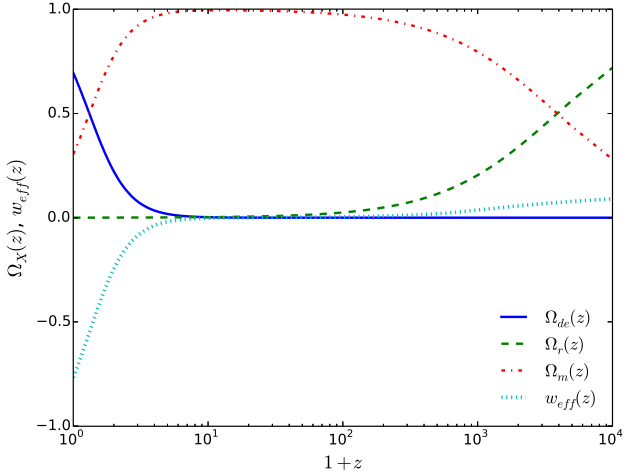}
\caption{The evolution of the dimensionless density parameters of $\Omega_{X=m,r,de}(z)$ and the effective EoS with respect to the redshift $z$ for the HS ($n=1$, $q_0=-0.65$ and $\log(|f_{R0}-1|)=-5.75$).}\label{fig:fROmegas}
\end{figure}
\end{center}

Taking the Hu-Sawicki (HS) model
\begin{equation}
f(R)=m^2\left[\frac{R}{m^2}- \frac{c_1 \left(\frac{R}{m^2}\right)^n}{1+c_2 \left(\frac{R}{m^2}\right)^n}\right],\label{Hu}
\end{equation}
as a working example in this paper, one obtains its first two derivatives of Eq. (\ref{Hu}) with respect to $R$ easily
\begin{eqnarray}
f_{R}&=&1-\frac{c_1  n \left(\frac{R}{m^2}\right)^{n-1}}{ \left[c_2 \left(\frac{R}{m^2}\right)^n+1\right]{^2}},\label{dHu1}\\
f_{RR}&=&\frac{1}{m^2}\frac{c_1 n \left(\frac{R}{m^2}\right)^{n-2} \left[c_2 (n+1) \left(\frac{R}{m^2}\right)^n-n+1\right]}{\left[c_2 \left(\frac{R}{m^2}\right)^n+1\right]{^3}}.\label{dHu2}
\end{eqnarray}
Only these three functions $f(R)$, $f_R$ and $f_{RR}$ are needed in our {\bf FRCAMB} package. In order to make the expansion history close to that of $\Lambda$CDM \cite{ref:HSmodel}, it was shown that the parameters $(c_1,c_2)$ were related to $\Omega_{m0}$, $\Omega_{r0}$
\begin{equation}
\frac{c_1}{c_2}=6\frac{ \left(1-\Omega_{r0}-
\Omega_{m0}\right)}{\Omega_{m0}}.
\end{equation}
Therefore for the HS model, we have three free model parameters $\{c_1,q_0,n\}$. In this work, we consider the $n=1$ case only. It is easy to extend to the $n\neq 1$ cases. We show the relative difference $(H_{f(R)}(z)-H_{GR}(z))/H_{GR}(z) \times 100 \%$ for the HS and $\Lambda$CDM model in Figure \ref{fig:fRH} by adopting the same cosmological parameters obtained by {\it Planck} \cite{ref:Planck2013CP} but with varying $c_1$ or effectively  varying $\log(|f_{R0}-1|)$. The results show that the background evolution of HS model for $n=1$ is very close to that of the $\Lambda$CDM. The relative deviation to the $\Lambda$CDM background is less than $0.07\%$. This also confirms the viability of our code for the background evolution. The curves for different values of $c_1$ imply that the background evolution is not sensitive to the values of $c_1$ for the HS model in  $n=1$ case. We also show the evolution of the dimensionless density parameter $\Omega_{X=m,r,de}(z)$ and the effective EoS with respect to the redshift $z$ in Figure \ref{fig:fROmegas}, where one can see a series of transitions of our Universe from the early radiation dominated epoch, the middle dark matter dominated epoch to the late effective dark energy dominated epoch. It says that the HS model pass the background evolution test.

Taking the trace of Eq. (\ref{eq:frEinstein}), one has
\begin{equation}
3\Box f_R-R+f_RR-2f=-\kappa^2(\rho_m+\rho_r).
\end{equation}
This equation can be recast as a evolution equation of a scalar field $f_R$
\begin{equation}
\Box f_R=\frac{\partial V_{eff}}{\partial f_R},
\end{equation}
with the effective potential
\begin{equation}
\frac{\partial V_{eff}}{\partial f_R}=\frac{1}{3}\left[R-f_R R+2f-\kappa^2(\rho_m+\rho_r)\right],
\end{equation}
which has an extremum at
\begin{equation}
R-f_R R+2f=\kappa^2(\rho_m+\rho_r).
\end{equation}
The effective mass of this scalar field at the extremum is given by
\begin{equation}
m^2_{f_{R}}=\frac{\partial^2 V_{eff}}{\partial f^2_R}=\frac{1}{3}\left(\frac{1+f_R}{f_{RR}}-R\right).
\end{equation}
Usually, instead of using the Compton wavelength $\lambda_{f_{R}}\equiv m^{-1}_{f_{R}}$, one defines the dimensionless Compton wavelength 
\begin{equation}
B=\frac{f_{RR}R^{\prime}}{1+f_{R}}\frac{H}{H^{\prime}}=\frac{f_{RR}R^{\prime}}{1+f_{R}}\frac{\frac{H^2}{m^2}}{\frac{1}{2}\left(\frac{H^2}{m^2}\right)^{\prime}}.
\end{equation}

\begin{center}
\begin{figure}[tbh]
\includegraphics[width=9.25cm]{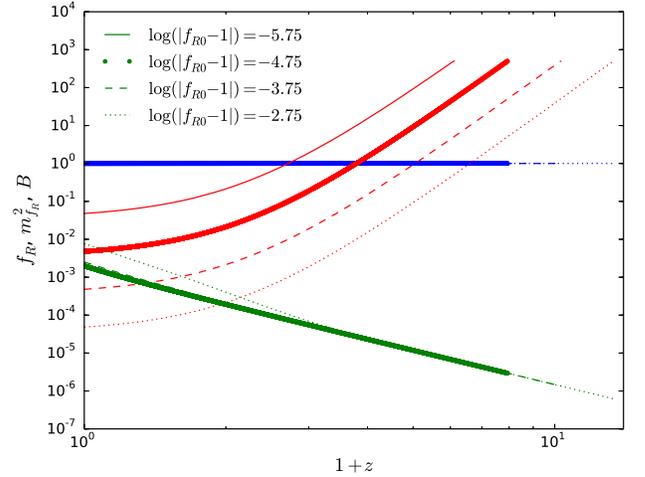}
\caption{The evolution of $f_{R}$ (blue curves), $m^2_{f_{R}}$ (red curves) and $B$ (green curves) with respect to the redshift $z$ for the Hu-Sawicki model $n=1$ with different values of $c_1$ or the effective $\log(|f_{R0}-1|)=-5.75,-4.75,-3.75,-2.75$. $m^2_{f_{R}}$ begins to increase quickly at a transition point $a_t$ dependent on the values of $c_1$ or the effective $\log(|f_{R0}-1|)=-5.75,-4.75,-3.75,-2.75$.}\label{fig:fRlog}
\end{figure}
\end{center}

Taking the HS model ($n=1$) as a working example, we show the evolution of $f_{R}$, $m^2_{f_{R}}$ and $B$ with respect to the redshift $z$ in Figure \ref{fig:fRlog}, where $m^2_{f_{R}}$ begins to increase quickly at a transition point dependent on the values of model parameter. It implies that the $f(R)$ gravity approaches to GR when the effective scalar field $f_{R}$ becomes massive. Therefore to make the code efficient and stable, we put a cutoff to the mass of $m^2_{f_{R}}\sim 5\times10^2 \text{ Mpc}^{-2}$ at the scale factor $a=a_{t}$ which value is determined by the model parameters. For the small values of $a<a_{t}$, we replace the evolution and perturbation with the standard GR model with the same cosmological model parameters. Actually in the early times $a<a_t$, the contribution of the effective dark energy component can be neglected as shown in Figure \ref{fig:fROmegas}.

\section{Perturbation Equations for an $f(R)$ Gravity Model} \label{sec:perturbation}

In this section, we will consider the scalar and tensor perturbation equations for an $f(R)$ gravity which has already been well studied in Ref. \cite{ref:frperturbationreview,ref:bean}. The line element with the scalar and tensor perturbation can be written as
\begin{eqnarray}
ds^2 &=&  a^2[-(1+2\Psi Y^{(s)})d\tau^2+2BY^{(s)}_id\tau
dx^i\nonumber \\
&+&(1+2\Phi Y^{(s)})\gamma_{ij}dx^idx^j+\mathcal{E}Y^{(s)}_{ij}dx^idx^j\nonumber\\
&+&2h_{T}Y^{(T)}_{ij}dx^{i}dx^{j} ]\label{pert_jordan},
\end{eqnarray}
where $\gamma_{ij}$ the three-dimensional spatial metric in the spherical coordinate is written as
\begin{equation}
[\gamma_{ij}]=\begin{pmatrix} \frac{1}{1-Kr^2} & 0 & 0 \\
               0 & r^2 & 0 \\
               0 & 0 & r^2\sin^2\theta
\end{pmatrix},
\end{equation}
and $Y^{(s)}$, $Y^{(s)}_{j}$, $Y^{(s)}_{ij}$ and $Y^{(T)}_{ij}$ are the scalar and tensor harmonic functions defined by
\begin{equation}
\begin{split}
&(\Delta+k^2)Y^{(s)}=0,\\
&Y_j^{(s)}\equiv-\frac{1}{k}Y_{|j}^{(s)},\\
&Y_{ij}^{(s)}\equiv\frac{1}{k^2}Y_{|ij}^{(s)}+\frac{1}{3}\gamma_{ij}Y^{(s)},\\
&(\Delta+k^2)Y_{ij}^{(T)}=0.
\end{split}
\end{equation}
In the synchronous gauge, by setting $\Psi=0$ and $B=0$ and 
\begin{eqnarray}
\eta_T&=&-(\Phi+\frac{\mathcal{E}}{6}),\\
h_L&=&6\Phi,
\end{eqnarray}
where $\eta_T$ refers to the conformal 3-space curvature perturbation
\begin{equation}
\delta R^{(3)}=6\delta K=-4(k^2-3K)\eta_T,
\end{equation}
the perturbed modified Einstein equations in the synchronous can be written as
\begin{eqnarray}
-\frac{1}{2}\kappa^2a^2\delta \rho&=&-(\frac{1}{2}F\mathcal{H}+\frac{1}{4}F')h_L'-\frac{3}{2}\mathcal{H}\delta F'-\frac{1}{2}\delta F k^2\nonumber \\
&+& F \eta_T(k^2-3K)+\frac{3}{2}\mathcal{H}'\delta F\label{SynE1}, \\
\kappa^2a^2\delta p&=&F[-\frac{2}{3}\mathcal{H}h_L'+\frac{2}{3}k^2\eta_T-\frac{1}{3}h_L''-2\eta_TK]\nonumber \\
&+&\delta F[\mathcal{H}^2+\frac{a''}{a}-\frac{2}{3}k^2+2K]-\frac{1}{3}F'h_L'\nonumber \\
&-&\delta F''-\delta F'\mathcal{H},\label{SynE2}\\
\alpha'&=&-2\mathcal{H}\alpha+\eta_T-\frac{F'}{F}\alpha -\kappa^2a^2\frac{p\Pi}{Fk^2}-\frac{\delta F}{F},\nonumber\\\label{SynE3}
\\
\frac{k^2-3K}{k}F\eta_T'&=&\frac{1}{2}\kappa^2a^2q+\frac{1}{2}k\delta F'-\frac{1}{2}k\mathcal{H}\delta F+\frac{Fh_L'K}{2k},\nonumber\\
\label{SynE4}
\end{eqnarray}
where
\begin{equation}
\begin{split}
\alpha&\equiv\frac{(h_L+6\eta_T)'}{2k^2},\\
      q&=(\rho+p)v,
\end{split}
\end{equation}
and
\begin{eqnarray}
&&\delta F''+2\mathcal{H}\delta F'+a^2(\frac{k^2}{a^2}+m_{f_{R}}^2)\delta F,\nonumber\\
 &=&\frac{\kappa^2a^2}{3}(\delta \rho -3\delta p) -\frac{1}{2}F'h_L'.\label{deltF}
\end{eqnarray}
In this section, we have used the notation $F\equiv1+f_{R}$ and the superscript $'=d/d\tau$. $\mathcal{H}=a'/a$ is the conformal Hubble parameter. 

In the {\bf CAMB} package, the curvature perturbations are characterized by $\mathcal{Z}$ and $\sigma$
\begin{equation}
\mathcal{Z}=\frac{h_L'}{2k}, \quad \sigma=k\alpha, \nonumber
\end{equation}
where
\begin{equation}
\eta_T'=\frac{k}{3}(\sigma-\mathcal{Z}).
\end{equation}
With the above variables, the perturbed modified Einstein equations recast into
\begin{eqnarray}
(F\mathcal{H}+\frac{1}{2}F')k\mathcal{Z}&=&\frac{\kappa^2}{2}a^2\delta \rho+Fk^2\eta_T\beta_2-\frac{3}{2}\mathcal{H}\delta F'\nonumber\\
&-&\frac{1}{2}\delta F k^2+\frac{3}{2}\mathcal{H}'\delta F,\\
\frac{k^2}{3}F(\beta_2\sigma-\mathcal{Z})&=&\frac{\kappa^2}{2}a^2q+\frac{1}{2}k\delta F'\nonumber \\
&-&\frac{1}{2}k\mathcal{H}\delta F,\\
\sigma'+2\mathcal{H}\sigma+\frac{F'}{F}\sigma&=&k\eta_T-\kappa^2a^2\frac{p\Pi}{F k}-k\frac{\delta F}{F},\\
\mathcal{Z}'+(\frac{1}{2}\frac{F'}{F}+\mathcal{H})\mathcal{Z}&=&(-k\beta_2+\frac{k}{2}+\frac{3\mathcal{H}^2}{k})\frac{\delta F}{F}\nonumber\\
     &-&\frac{\kappa^2a^2}{2kF}(\delta \rho+3\delta p)-\frac{3}{2}\frac{\delta F''}{kF},
\end{eqnarray}
where
\begin{equation}
\beta_2=\frac{k^2-3K}{k^2},
\end{equation}
is the curvature factor. The propagation of the perturbed field $\delta F$ is given by
\begin{eqnarray}
&&\delta F''+2\mathcal{H}\delta F'+a^2(\frac{k^2}{a^2}+m_{f_{R}}^2)\delta F\nonumber\\
 &=&\frac{\kappa^2a^2}{3}(\delta \rho -3\delta p) -kF'\mathcal{Z}.\label{deltF}
\end{eqnarray}

The source term of the CMB temperature anisotropy is given by \cite{ref:source,ref:cambequations}
\begin{equation}
\begin{split}
&S_T(\tau,k)\\
&=e^{-\varepsilon}(\alpha''+\eta_T')\\
&+g(\Delta_{T0}+2\alpha'+\frac{v_b'}{k}+\frac{\zeta}{12\sqrt{\beta_2}}+\frac{\zeta''}{4k^2\sqrt{\beta_2}})\\
&+g'(\alpha+\frac{v_b}{k}+\frac{\zeta'}{2k^2\sqrt{\beta_2}})+\frac{1}{4}\frac{g''\zeta}{k^2\sqrt{\beta_2}}\\
&=e^{-\varepsilon}(\frac{\sigma''}{k}+\frac{k\sigma}{3}-\frac{k\mathcal{Z}}{3})\\
&+g(\Delta_{T0}+2\frac{\sigma'}{k}+\frac{v_b'}{k}+\frac{\zeta}{12\sqrt{\beta_2}}+\frac{\zeta''}{4k^2\sqrt{\beta_2}})\\
&+g'(\frac{\sigma}{k}+\frac{v_b}{k}+\frac{\zeta'}{2k^2\sqrt{\beta_2}})+\frac{1}{4}\frac{g''\zeta}{k^2\sqrt{\beta_2}},
\end{split}
\end{equation}
where $g=-\dot{\varepsilon}e^{-\varepsilon}=an_e\sigma_Te^{-\varepsilon}$ is the visibility function and $\varepsilon$ is the optical depth. $\zeta$ is given by
\begin{equation}
\zeta=(\frac{3}{4}I_2 +\frac{9}{2}E_2),
\end{equation}
where $I_2$, $E_2$ indicate the quadrupole of the photon intensity and the E-like polarization respectively \cite{ref:cambequations}.

The propagation of gravitational waves for an $f(R)$ gravity is given by
\begin{equation}
h''_{T}+2\mathcal{H}\left(1+\frac{1}{2}\frac{d\ln F}{d\ln a}\right)h'_{T}+c^2_T(k^2+2K)h_{T}=\frac{8\pi G a^2 \Pi}{F},\label{eq:FRtensor}
\end{equation}
where $c^2_T$ is the square of the speed of gravitational waves.

For the perturbation equations, one can clearly see that the only $f(R)$ gravity model dependent term is $m^2_{f_{R}}$ in the perturbed field equation (\ref{deltF}) for $\delta F$. For a concrete $f(R)$ form, $m^2_{f_{R}}$ can be calculated easily from its first two derivatives of $f(R)$ with respect to $R$. 

\section{CMB Power Spectrum and Matter Power Spectrum}\label{sec:powers}

In this section, we will show the CMB power spectrum and the linear matter power spectrum for an $f(R)$ gravity model. As mentioned in the previous section, the only inputs for an $f(R)$ gravity model are the three functions $f(R)$, $f_{R}\equiv df(R)/dR$ and $f_{RR}\equiv d^2f(R)/dR^2$. By setting the proper values of the model parameter for an $f(R)$ gravity model, our code {\bf FRCAMB} will calculate the background evolution, solve the Einstein-Boltzmann equation and output the CMB power spectrum, the matter power spectrum, the effective EoS $w_{eff}(a)$ and the dimensionless energy density $\Omega_{X}(a)$, and almost all the quantities you are interested in.    

\begin{center}
\begin{figure}[tbh]
\includegraphics[width=9.25cm]{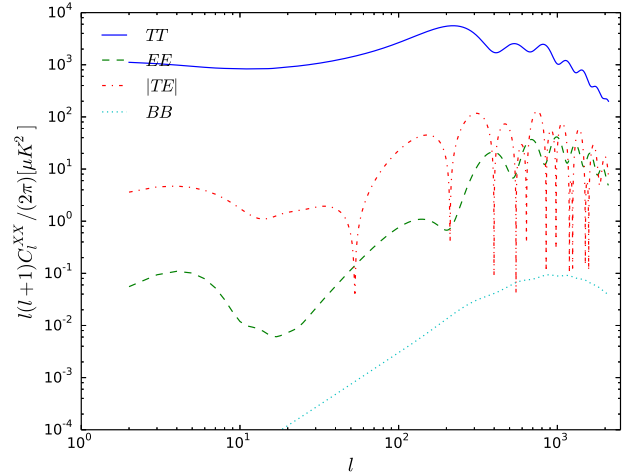}
\caption{The CMB TT, EE, TE and BB power spectrum for HS model with $n=1$, $\log(|f_{R0}-1|)=-5.75$.}\label{fig:CMBXX}
\end{figure}
\end{center}

\begin{center}
\begin{figure}[tbh]
\includegraphics[width=9.25cm]{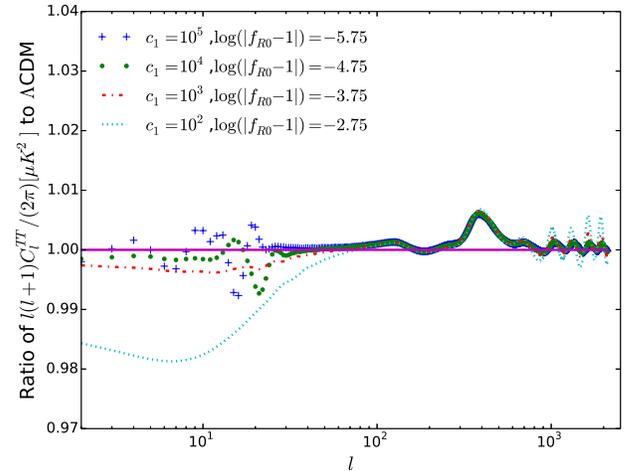}
\caption{The ratio of the CMB TT power spectrum for different values of $\log(|f_{R0}|)$ to that of the $\Lambda$CDM cosmology with the same values of the cosmological parameters.}\label{fig:CMB}
\end{figure}
\end{center}

\begin{center}
\begin{figure}[tbh]
\includegraphics[width=9.25cm]{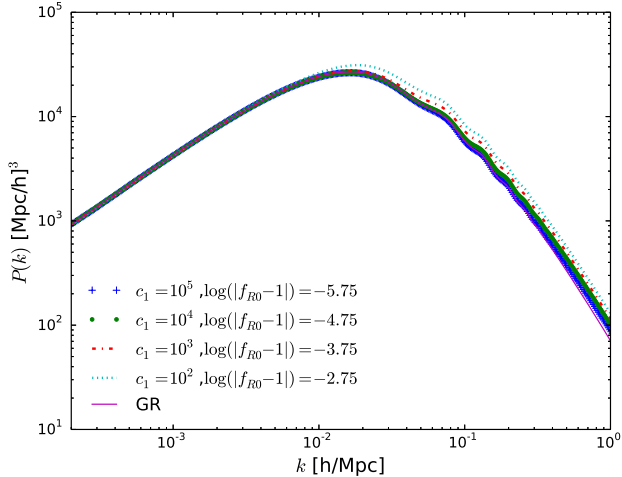}
\caption{The linear matter power spectrum at the redshift $z=0$ for different values of $\log(|f_{R0}-1|)$ and GR with the same values of the cosmological parameters.}\label{fig:MP}
\end{figure}
\end{center}

Taking the HS model ($n=1$) as a working example, we show the CMB $XX=TT,EE,TE,BB$ power spectrum in Figure \ref{fig:CMBXX}, the effects on the CMB TT power spectrum and the linear matter power spectrum for different values of the model parameter $c_1$ or the effective $\log(|f_{R0}-1|)$ in Figure \ref{fig:CMB} and Figure \ref{fig:MP}, where $q_0=-0.65$ is fixed for illustration.

As shown in Figure \ref{fig:CMB}, the CMB TT power spectrum is much sensitive to the values of the model parameter $c_1$ or its equivalent $\log(|f_{R0}-1|)$ at low $\ell$ multipole where it is dominated by the late integrated Sachs-Wolfe (ISW) effect. This is mainly due to the fact that at the early epoch of our Universe the mass of the effective scalar field $f_{R}$ becomes massive and there is no significant deviation to $\Lambda$CDM model, because $\Lambda$CDM cosmology is switched on when $m^2_{f_{R}}\sim 5\times 10^2 \text{ Mpc}^{-2}$ is arrived as shown in Figure \ref{fig:fRlog}. The same effects can be seen in Figure \ref{fig:MP}, the linear matter power spectrum is sensitive to the values of $c_1$ or its equivalent $\log(|f_{R0}-1|)$ after the matter-radiation equality epoch. 

Once we have the linear matter power spectrum, the nonlinear matter power spectrum can be calculated through a routine like {\bf HALOFIT} \cite{ref:HALOFIT}. For this kind of halo-fit formula for an $f(R)$ gravity model, an $N$-body simulation is strongly demanded. We already have {\bf MGHalofit} for calculating the nonlinear matter power spectrum \cite{ref:MGhalofit}, but it is only for the HS model ($n=1$) and works in the range $|f_{R0}|\in[10^{-6},10^{-4}]$ and $z\le 1$. A general halo-fit formula for any $f(R)$ gravity is still unavailable. In our previous study on a specific family of $f(R)$ gravity model, it is found out that the redshift space distortion (RSD) $f\sigma_8$ can provide a tight constraint to the values of $|f_{R0}-1|\sim 10^{-6}$ \cite{ref:Xufr}. For this small values of $|f_{R0}-1|\sim 10^{-6}$, the nonlinear matter power spectrum can almost mimic that of the $\Lambda$CDM model. With the very small values of $f_{R0}$, it is difficult to detect a model not only because of the accuracy of the fitting formula but also the complicated astrophysical systematics on such scales \cite{ref:MGhalofit}.  

\section{Constraint to HS model from Cosmic Observations}

In this section, we show the constraint results to HS model for the $n=1$ case from the geometric and dynamic measurements. For the geometrical one, we will use the supernova Ia data from SDSS-II/SNLS3 joint light-curve analysis \cite{ref:SNJLA}, the baryon acoustic oscillation $D_V(0.106) = 456\pm 27$ [Mpc] from 6dF Galaxy Redshift Survey \cite{ref:BAO6dF}; $D_V(0.35)/r_s = 8.88\pm 0.17$ from SDSS DR7 data \cite{ref:BAOsdssdr7}; $D_V(0.57)/r_s = 13.62\pm 0.22$ from BOSS DR9 data \cite{ref:sdssdr9}, the present Hubble parameter $H_0 = 73.8\pm 2.4$ [$\text{km s}^{-1} \text{Mpc}^{-1}$] from HST \cite{ref:HST}, and the full information of CMB recently released by {\it Planck}2013 (which include the high-l TT likelihood ({\it CAMSpec}) up to a maximum multipole number of $l_{max}=2500$ from $l=50$, the low-l TT likelihood ({\it lowl}) up to $l=49$) \cite{ref:Planckdata} with the addition of the low-l TE, EE, BB likelihood up to $l=32$ from WMAP9. For the dynamical one, we use the redshift space distortion (RSD) data. For using the growth rate, we calculate the $f\sigma_8(z)=d\sigma_8/d\ln a$ at different redshifts in theory. For details, please see Ref. \cite{ref:Xufr}.

\begin{center}
\begin{figure}[tbh]
\includegraphics[width=9.25cm]{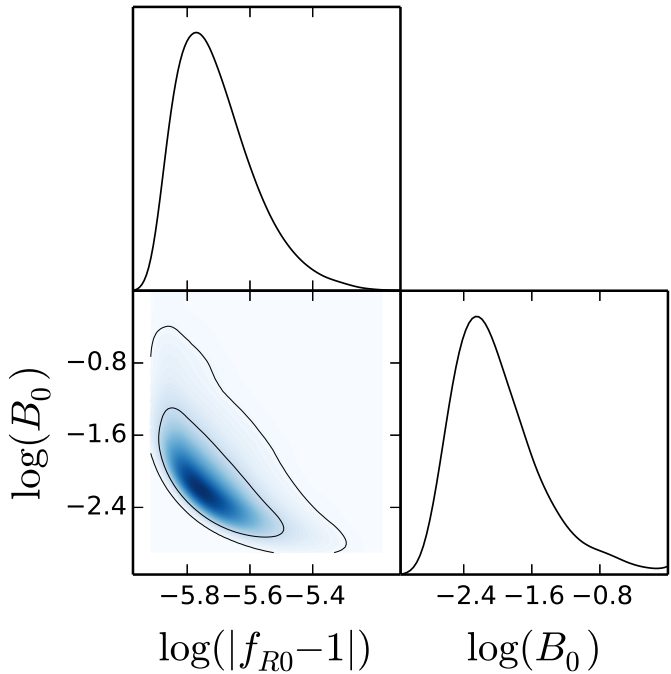}
\caption{The 1D marginalized distribution and 2D contours for interested model parameters with $68\%$ C.L., $95\%$ C.L. by using the {\it Planck} 2013, WMAP9, BAO, BAO, JLA, HST and RSD data sets.}\label{fig:contour}
\end{figure}
\end{center}

We perform a global fitting on the {\it Computing Cluster for Cosmos} by using the publicly available package {\bf CosmoMC} \cite{ref:MCMC} in the following model parameter space
\begin{equation}
P=\{\Omega_b h^2,\Omega_c h^2,  100\theta_{MC}, \tau, n_s, {\rm{ln}}(10^{10} A_s),q_0, \log(c_1)\},
\end{equation}
their priors are shown in the second column of Table \ref{tab:results}. The running was stopped when the Gelman \& Rubin $R-1$ parameter $R-1 \sim 0.02$ was arrived; that guarantees the accurate confidence limits. The obtained results are shown in Table \ref{tab:results} and Figure \ref{fig:contour}. 
\begingroup                                                                                                                     
\begin{center}                                                                                                                  
\begin{table}                                                                                                                   
\begin{tabular}{cccc}                                                                                                            
\hline\hline                                                                                                                    
Parameters & Priors & Mean with errors & Best fit \\ \hline
$\Omega_b h^2$ & $[0.005,0.1]$  & $0.02242_{-0.00026}^{+0.00025}$ & $0.02241$\\
$\Omega_c h^2$ & $[0.001,0.99]$  & $0.1160_{-0.0015}^{+0.0014}$ & $0.1167$\\
$100\theta_{MC}$ & $[0.5,10]$ & $1.04169_{-0.00056}^{+0.0005}$ & $1.04175$\\
$\tau$ & $[0.01,0.81]$ & $0.074_{-0.011}^{+0.011}$ & $0.071$\\
$q_0$ & $[-1,0]$ &  $-0.787_{-0.21}^{+0.046}$ & $-0.920$\\
$\log(c_1)$ & $[0,5]$ & $4.906_{-0.019}^{+0.094}$ & $4.994$\\
${\rm{ln}}(10^{10} A_s)$ & $[2.7,4]$ & $3.046_{-0.021}^{+0.021}$ & $3.042$\\
$n_s$ & $[0.9,1.1]$ & $0.9678_{-0.0056}^{+0.0055}$ & $0.9664$\\
\hline
$H_0$ & $...$ & $69.03_{-0.68}^{+0.68}$ & $68.81$\\
$\Omega_{DE}$ & $...$ & $0.7080_{-0.0082}^{+0.0083}$ & $0.7049$\\
$\Omega_m$ & $...$ & $0.2920_{-0.0083}^{+0.0082}$ & $0.2951$\\
$\sigma_8$ & $...$ & $0.8214_{-0.0098}^{+0.0100}$ & $0.8168$\\
$z_{\rm re}$ & $...$ & $9.48_{-0.96}^{+0.99}$ & $9.24$\\
${\rm{Age}}/{\rm{Gyr}}$ & $...$ & $13.752_{-0.037}^{+0.037}$ & $13.754$\\
$\log(|f_{R0}-1|)$ & $...$ & $-5.71_{-0.16}^{+0.07}$ & $-5.86$\\
$\log(B_0)$ & $...$ & $-1.99_{-0.63}^{+0.28}$ & $-1.78$\\
\hline\hline                                                                                                                    
\end{tabular}                                                                                                                   
\caption{The mean and best fit values with $1\sigma$ errors for the interested and derived cosmological parameters, where the {\it Planck} 2013, WMAP9, BAO, SN, HST and RSD data sets were used.}\label{tab:results}                                                                                                
\end{table}                                                                                                                     
\end{center}                                                                                                                    
\endgroup  

\section{Conclusion} \label{sec:conclusion} 

In this paper, we present an Einstein-Boltzmann equation solver, named {\bf FRCAMB}, for calculating the anisotropies in the microwave background in any $f(R)$ gravity model based on a modified version of {\bf CAMB}. In this code, instead of assuming a $\Lambda$CDM or $w$CDM background cosmology as done in the {\bf EFTCAMB} code and the {\bf MGCAMB} code, we solve the background evolution numerically for any $f(R)$ gravity model once the $f(R)$ function, its first two derivative with respect to $R$, i.e. $f_R\equiv df/dR$, $f_{RR}\equiv d^2f/dR^2$ and the reasonable values of the $f(R)$ model parameters are inputted. The outputs of this code include the CMB power spectrum, the matter power spectrum, the evolution of the total effective EoS of our Universe and almost everything interesting. By global fitting to the model parameter space through the geometrical and dynamical cosmic observations, we obtain $\log(|f_{R0}-1|)=-5.71_{-0.16}^{+0.07}$ which is consistent to the previous result obtained by {\bf MGCAMB} and {\bf EFTCAMB}.

\acknowledgements{The author thanks Prof. A. A. Sen, M. Raveri and Dr. Y. Wang for useful discussion and ICTP for hospitality during the author's visit in ICTP. This work is supported in part by National Natural Science Foundation of China under Grant No. 11275035 (People's Republic of China), and the Open Project Program of State Key Laboratory of Theoretical Physics, Institute of Theoretical Physics, Chinese Academy of Sciences No. Y4KF101CJ1 (People's Republic of China).}


\begin{thebibliography}{*}

\bibitem{ref:book} L. Amendola, S. Tsujikawa, Cambridge University Press, 2010.

\bibitem{ref:planck2015MG} Planck Collaboration: P. A. R. Ade, et al, arXiv:1502.01590 [astro-ph.CO].

\bibitem{ref:FRCAMB} J.-h. He, Phys. Rev. D86, 103505 (2012), http://darklight.brera.inaf.it/cosmonews/frcamb/.

\bibitem{ref:MGCAMB} A. Hojjati, L. Pogosian, G.-B. Zhao, JCAP 1108,005 (2011), arXiv:1106.4543 [astro-ph.CO], http://www.sfu.ca/~aha25/MGCAMB.html.

\bibitem{ref:EFTCAMB} B. Hu, M. Raveri, A. Silvestri, N. Frusciante, Phys. Rev. D 91, 063524 (2015), arXiv:1410.5807 [astro-ph.CO], http://wwwhome.lorentz.leidenuniv.nl/~hu/codes/.

\bibitem{ref:CAMB} A. Lewis, A. Challinor, A. Lasenby, Astrophys. J. 538, 473 (2000), arXiv:astro-ph/9911177, http://camb.info

\bibitem{ref:MGR1} A. Silvestri and M. Trodden, Rept. Prog. Phys. 72, 096901 (2009) [arXiv:0904.0024 [astro-ph.CO]].

\bibitem{ref:MGR2} T. Clifton, P. G. Ferreira, A. Padilla and C. Skordis, arXiv:1106.2476 [astro-ph.CO].

\bibitem{ref:MGR3} A. Joyce, B. Jain, J. Khoury, M. Trodden, arXiv:1407.0059 [astro-ph.CO]. 

\bibitem{ref:MGR4} S. Nojiri, S. D. Odintsov, Phys. Rept. 505, 59(2011).    


\bibitem{ref:HSmodel} W. Hu, I. Sawicki, Phys. Rev. D 76, 064004 (2007).

\bibitem{ref:Planck2013CP} P. A. R. Ade, et al, (Planck Collaboration), arXiv:1303.5076 [astro-ph.CO]. 

\bibitem{ref:frperturbationreview} J.-c. Hwang, H. Noh, Phys. Rev. D 65 (2001) 023512; J.-c. Hwang, H. Noh, Phys. Rev. D, 71 (2005) 063536.   

\bibitem{ref:bean} R. Bean, D. Bernat, L. Pogosian, A. Silvestri, M. Trodden, Phys.Rev.D75 (2007) 064020, astro-ph/0611321.


\bibitem{ref:source} M. Zaldarriaga, U. Seljak and E. Bertschinger, Astrophys. J. 494 (1998) 491, astro-ph/9704265;

\bibitem{ref:cambequations} A. Challinor, Phys.Rev. D62 (2000) 043004, astro-ph/9911481.


\bibitem{ref:HALOFIT} Smith, R. E., Peacock, J. A., Jenkins, A., et al., MNRAS,
341(2003)1311 (S03); R. Takahashi, M. Sato, T. Nishimichi, A. Taruya, M. Oguri, ApJ, 761(2012)152, arXiv:1208.2701 [astro-ph.CO]. 

\bibitem{ref:MGhalofit} G.-B. Zhao, ApJS, 211, 23 (2014), arXiv:1312.1291 [astro-ph.CO]. 

\bibitem{ref:Xufr} L. Xu, Phys. Rev. D 91, 063008 (2015), arXiv:1411.4353 [astro-ph.CO].   


\bibitem{ref:SNJLA} M. Betoule, et al., arXiv:1401.4064 [astro-ph.CO], http://supernovae.in2p3.fr/sdss\_snls\_jla/ReadMe.html.

\bibitem{ref:BAO6dF} F. Beutler, et al., Mon. Not. Roy. Astron. Soc. 416, 3017 (2011), arXiv:1106.3366 [astro-ph.CO].

\bibitem{ref:BAOsdssdr7} N. Padmanabhan, et al., Mon. Not. Roy. Astron. Soc. 427, 2132 (2012), arXiv:1202.0090 [astro-ph.CO].

\bibitem{ref:sdssdr9} L. Anderson, et al., Mon. Not. Roy. Astron. Soc. 428, 1036 (2013) arXiv:1203.6594 [astro-ph.CO].

\bibitem{ref:HST} A. G. Riess, et al., ApJ, 730, 119 (2011), arXiv:1103.2976[astro-ph.CO].

\bibitem{ref:Planckdata} P. A. R. Ade, et al, (Planck Collaboration), arXiv:1303.5076 [astro-ph.CO], http://pla.esac.esa.int/pla/aio/planckProducts.html.

\bibitem{ref:MCMC} A. Lewis and S. Bridle, Phys. Rev. D 66, 103511 (2002); http://cosmologist.info/cosmomc/.

\end{thebibliography}
\end{document}